\RequirePackage{lineno}
\documentclass[twocolumn,a4paper,prstab,showkeys,showpacs,superscriptaddress]{revtex4}

\usepackage{amsmath}    % need for subequations
\usepackage{graphicx}   % need for figures
\usepackage{verbatim}   % useful for program listings
\usepackage{color}      % use if color is used in text
\usepackage{subfigure}  % use for side-by-side figures
\usepackage{hyperref}   % use for hypertext links, including those to external documents and URLs
\usepackage{multirow}
\usepackage{dcolumn}		% Align table columns on decimal point
\usepackage{bm}					% bold math

\begin{document}
%\linenumbers

\title{Observation of the Stimulated Coherent Diffraction Radiation in an Open Resonator at LUCX Facility}

\author{A. Aryshev}
 \email{alar@post.kek.jp} 
\author{S. Araki}
\author{M. Fukuda}
\author{N. Terunuma}
\author{J. Urakawa}
\affiliation{KEK: High Energy Accelerator Research Organization,1-1 Oho, Tsukuba, Ibaraki 305-0801, Japan}
\author{P. Karataev}
\affiliation{John Adams Institute at Royal Holloway, University of London, Egham, Surrey, TW20 0EX, United Kingdom}
\author{G. Naumenko}
\author{A. Potylitsyn}
\author{L. Sukhikh}
\author{D. Verigin}
\affiliation{Tomsk Polytechnic University, Institute of Physics and Technology, Lenin ave. 30, Tomsk 634050, Russian Federation}
\author{K. Sakaue}
\affiliation{Waseda University, 3-4-1, Okubo, Shinjuku, Tokyo, 169-8555, Japan}

\begin{abstract} We present an initial test of a new type of a pre-bunched beam pumped free electron maser based on Stimulated Coherent Diffraction Radiation (SCDR) generated in an open resonator. An ultra-fast Schottky Barrier Diode (time response $<1\,$ns) has enabled to investigate the properties of the radiation stored in the cavity as well as the intrinsic properties of the cavity itself. We observed a turn-by-turn SCDR generated by a multibunch beam. When the cavity length was exactly a half of the bunch spacing a clear resonance was observed. Moreover, turn-by-turn measurements revealed the cavity quality factor of $72.88$, which was rather high for an open resonator in the wavelength range of $3-5\,$mm. An exponential growth of the photon intensity as a function of the number of bunches was also demonstrated.
\end{abstract}


\maketitle

\section{Introduction}

Electromagnetic radiation in the frequency range from $100\,$GHz to a few THz has intensively been used for research and development in physics, material science, chemistry, biology and medicine. A very wide range of applications includes technologies and techniques for monitoring and control of charged particle beams, home land security, spectroscopy of superconductors and imaging of cancerous cells due to the fact that small molecules oscillate at THz frequencies. At the Workshop on Opportunities in THz Science organized by THz Science$\&$Technology Network \cite{1} it was concluded that scientific community requires a few technical needs: high peak fields and coverage to $10\,$THz (or higher) by coherent broad-band sources (which can only be achieved in particle accelerators); excellent stability of the sources; full pulse shaping; sensitive non-cryogenic detectors and easy access to THz components.

The process of stimulated radiation is a basis for laser physics, including traditional Free Electron Lasers (FEL), where electrons moving on a periodical path (for example, in an undulator) are accompanied by an electromagnetic wave, which wavelength is defined by the undulator period and the particle Lorentz-factor \cite{2,3}. In this case, the initial bunch is modulated longitudinally due to its interaction with the generated wave and an enhancement of the resultant radiation (or Stimulation) occurs. The modulated filaments of the beam might generate radiation coherently for wavelengths much shorter than overall bunch length.

Recent experimental observations of Stimulated Transition Radiation (STR) \cite{4} and Stimulated Synchrotron Radiation (SSR) \cite{5} show that STR and SSR have continuous spectra in contrast to the undulator radiation. Also the concept of a so-called ``broadband FEL'' based on STR has been proposed in \cite{5}. In comparison with traditional FELs, where radiation of an accelerated charge is stimulated along its entire trajectory, in STR experiments \cite{5,6,7} the radiation generated from target surfaces interacts with  the pre-bunched electron beam. To ignite the stimulation process it is necessary to apply high external EM field strength at the radiating surface. One of the possible ways to do that is to use a microwave resonator where a reasonable EM field strength can be accumulated. In this case radiation emission is \textit{stimulated} by the presence of an external field \cite{8a}. For that purpose an ``open resonator'' \cite{6} and a ``closed resonator'' \cite{7} have been used. In above mentioned schemes, authors have demonstrated their ability to generate intense coherent THz radiation with broad spectrum, however, its use for developing  a radiation source on a basis of a compact moderate relativistic energy Linacs with a high population beam is very limited due to the following reasons:

\begin{itemize}
\item  the properties of the resonator mirrors might change due to direct impact of the high intensity beam;
\item  interaction of a low energy beam with target material might significantly change beam parameters and affect the stimulation process, changing the STR properties;
\item  direct interaction of a high intensity beam (even with a moderate relativistic energy) with the target material will result in generation of a high intensity gamma- and X-ray radiation background which violates the radiation protection requirements in all modern facilities.
\end{itemize}

These facts might significantly limit the source performance. However, all these limitations might be eliminated if the broadband FEL is based on Coherent Diffraction Radiation (CDR). In this case the beam propagates through axial openings in resonator mirrors. If the opening radii satisfy the condition $r_{hole} \ll \gamma\lambda$, where $\gamma$  is the charged particle Lorentz factor and $\lambda$ is the radiation wavelength, then the CDR characteristics nearly coincide with CTR ones \cite{8}. While the bunch train passes through the resonator, it will ``accumulate'' the radiation power from many bunches. Moreover, every bunch will generate radiation in presence of the field generated by the preceding bunch stimulating the CDR emission. This radiation we call as \textit{Stimulated Coherent Diffraction Radiation} (SCDR).

It was decided to investigate the SCDR process as a potential candidate for generating intense broad-band radiation in THz and sub-THz frequency range as a part of a larger THz program launched at KEK: LUCX (Laser Undulator Compact X-ray source) facility. The program is aiming to investigate various mechanisms for generating EM radiation including SCDR, Undulator radiation, Smith-Purcell and other special cases of Polarization Radiation. In this paper we overview the initial test of a pre-bunched beam pumped free electron maser based on SCDR. The properties of the stored radiation, quality factor of the open resonator via direct observation of the stored radiation using an ultra-fast room-temperature Schottky Barrier Diode detector were investigated and alignment accuracies for optimal performance of the device were studied.

\section{Theoretical Background}

\subsection{Coherent radiation}

Coherent radiation is generated when the wavelength of the emitted radiation is comparable to or longer than the charged particles bunch length, i.e. emitted radiation fields of all electrons will add up coherently since they are emitted at roughly the same time. The expression for radiation spectrum $\frac{dW}{d\omega }$ can be represented as a sum of coherent and incoherent parts \cite{9}:
\begin{equation} \label{GrindEQ__1_} 
\frac{dW}{d\omega } =\frac{dW_{e} }{d\omega } \, \left[N+N\left(N-1\right)\left|F\left(\omega \right)\right|^{2} \right].     
\end{equation} 
Here $\frac{dW_{e} }{d\omega } $ is the radiation spectrum generated by a single electron, $N$ is the bunch population, and $F\left(\omega \right)$ is the longitudinal bunch form factor which is defined as a Fourier transform of the longitudinal particle distribution in a bunch, $\rho\left(\omega \right)$ \cite{8}. If the bunch has a Gaussian longitudinal profile, then:
\begin{equation} \label{GrindEQ__2_} 
F\left(\omega \right)=\int _{-\infty }^{\infty }\rho \left(z\right)e^{i\frac{\omega }{c\beta } z} dz =e^{-\frac{\omega ^{2} \sigma _{z}^{2} }{2c^{2} \beta ^{2} } }  
\end{equation} 
Here $z$ is the longitudinal coordinate, $c$ is the speed of light, $\beta$ is the velocity of a particle in units of the speed of light, $\omega$ is the frequency of light, and $\sigma_{z}$ is the RMS bunch length.

One may see that the first term in Eq.\eqref{GrindEQ__1_} is responsible for the incoherent part of the spectrum and proportional to the number of particles. The second part is responsible for the coherent radiation emission and proportional to the number of particles squared. As a result, the form factor in Eq.\eqref{GrindEQ__2_} tends to unity leading to a significant increase in intensity which is quadratic versus the bunch population. 

To be able to generate and precisely control THz and sub-THz radiation beams it is necessary to generate and fully characterize extremely short ($\sigma_{z} < 100\,$fs) high intensity beam of electrons. The LUCX facility described in section \ref{sES} aims to achieve that goal.

\subsection{CDR spatial distribution}

Nowadays the theory for laser cavities in optical and near infrared spectral ranges is very well developed. However, the theory for the wavelengths comparable to the dimensions of the cavity mirrors is rather complicated as it is necessary to take into account severe diffraction effects. The simulation of every reflection in this case will lead to a double integration over the mirror surfaces. Even a few round-trips will require an outstanding computation power. That is why the theoretical investigation of a complete power damping in the cavity becomes impractical. Another possibility is to investigate the SCDR properties with one of the simulation tools, such as CST Studio Suite \cite{10}, which is a separate complicated work which requires significant time. A comparison of the experimental THz power generation in a open resonator with simulation code is already in the authors plan and will be accomplished in a dedicated work.

\begin{figure}[htbp!]
\includegraphics{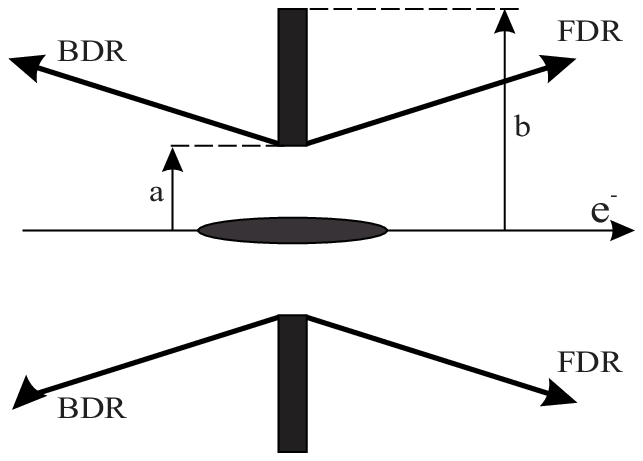}
\caption{DR calculation geometry.}
\label{F1}
\end{figure}

Nevertheless, many characteristics could be understood from an experiment. To be able to obtain the spatial characteristics of the radiation generated and accumulated within an open resonator it is necessary to know the CDR spatial characteristics produced by the beam from a single mirror of the resonator. When a charged particle passes in the vicinity of a conducting surface of a mirror (e.g. moves through a hole in a screen) the generated DR propagates in two directions: along the particle trajectory -- Forward Diffraction Radiation (FDR) and in the direction of specular reflection -- Backward Diffraction Radiation (BDR), as it is shown in Fig.\ref{F1}. It is important to point out that for an ideally conducting screen BDR and FDR characteristics coincide \cite{8}. For THz frequencies the approximation of an ideal conductor is very well suited. 

From Eq.\eqref{GrindEQ__1_} it is clear that if the bunch form factor is equal to unity the entire bunch acts as a single particle of enormous charge. Therefore the expression for diffraction radiation generated by a single electron will give us an idea about the CDR spatial characteristics. In accordance with \cite{11} the DR spectral-angular distribution generated by a single electron crossing a disc of radius $b$ with a hole of radius $a$ (see Fig.\ref{F1}) can be represented in the following form:

\begin{widetext}

\begin{equation} \label{GrindEQ__3_} 
\frac{d^{2}W_{e}}{\hbar d\omega d\Omega}=\frac{2\alpha }{\pi^{2}} \frac{\theta^{2}}{\left(\theta^{2}+\gamma^{-2}\right)^{2}}\left[\frac{2\pi fb}{\gamma c} K_{1} \left(\frac{2\pi fb}{\gamma c} \right) J_{0} \left(\frac{2\pi fb}{c} \theta \right) - \frac{2\pi fa}{\gamma c} K_{1} \left(\frac{2\pi fa}{\gamma c} \right) J_{0} \left(\frac{2\pi fa}{c} \theta \right)\right]^{2}  
\end{equation} 

\end{widetext}

Here $\alpha$ is the fine structure constant; $\theta$ is the polar observation angle; $b$ is the outer radius of the mirror and $a$ is the radius of the hole in the mirror; and $f = c/\lambda$ is the radiation frequency.

One may see that the expression is independent of the azimuthal angle of the photon emission, i.e. the distribution is azimuthally symmetric.  However, the distribution can always be represented as a sum of two orthogonal polarization components \cite{8}. Previously it has been demonstrated \cite{12} that the finite outer target dimensions result in significant change of the DR spatial distribution and suppression of the DR photon yield at longer wavelengths. To be able to assess the spatial properties of the radiation stored in the resonator one should integrate Eq.\eqref{GrindEQ__3_} over the wavelength response efficiency of the detector in the following form:

\begin{eqnarray} \label{GrindEQ__4_} 
\frac{dW_{e} }{d\Omega } & = &\int _{\hbar \omega _{\min } }^{\hbar \omega _{\max } }\frac{d^{2} W_{e} }{\hbar d\omega d\Omega } \cdot \hbar d\omega = \\
& = & \int _{\lambda _{\min } }^{\lambda _{\max } }\frac{d^{2} W_{e} }{\hbar d\omega d\Omega } \cdot \frac{2\pi \hbar c}{\lambda ^{2} } d\lambda    \nonumber
\end{eqnarray} 
where $\lambda_{min}=3.33\,$mm and $\lambda_{max}=5\,$mm. Here uniform sensitivity of the detector across the wavelength range is assumed.

\begin{figure}
\includegraphics{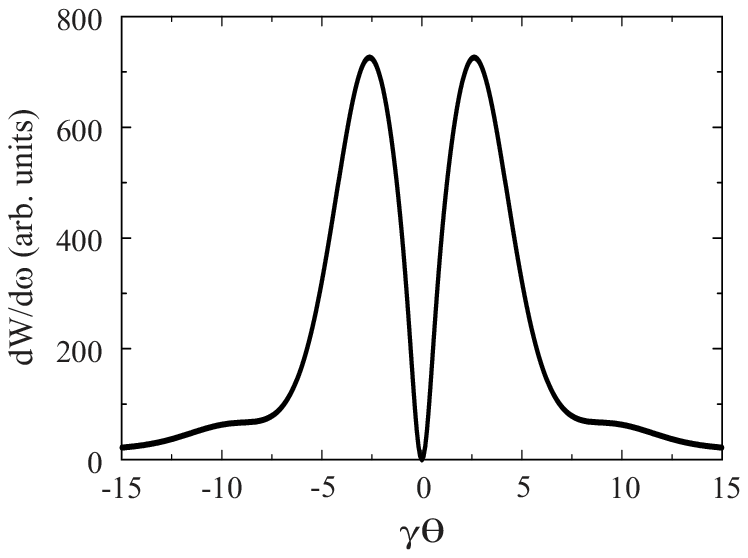}
\caption{CDR spatial distribution calculated using Eq. \eqref{GrindEQ__4_} for $b = 50\,$mm, $a = 3\,$mm, $\gamma = 80$.}
\label{F2}
\end{figure}

The result of the calculation is illustrated in Fig.\ref{F2}. One should note that this model does not take into account the pre-wave zone effect \cite{13} and the diffraction effect when the light propagates through the output vacuum viewport \cite{14}. Both effects result in broadening of the DR angular distribution. Nevertheless, it is clear that the DR distribution has two maxima and a minimum in the centre. By measuring the DR orientation dependence, i.e. the dependence of the photon yield as a function of the mirror angle, it is possible to find back reflection and align two resonator mirrors with respect to each other.

Another important characteristic is the DR spectral content. This experimental work is devoted to development of a new methods for generating THz radiation. The following estimations enable us to show how the spectrum extends to THz frequencies. The DR spectrum from a single electron can be calculated in the following form:

\begin{eqnarray} \label{Eq5} 
\frac{dW_{e} }{d\omega } & = & \int _{0}^{2\pi }\int _{0}^{\theta _{\max } }\frac{d^{2} W_{e} }{d\omega d\Omega } \cdot \theta d\theta d\phi = \\
& = & 2\pi \int _{0}^{\theta _{\max } }\frac{d^{2} W_{e} }{d\omega d\Omega } \cdot \theta d\theta   \nonumber
\end{eqnarray} 

\begin{figure}[htp!]
\includegraphics{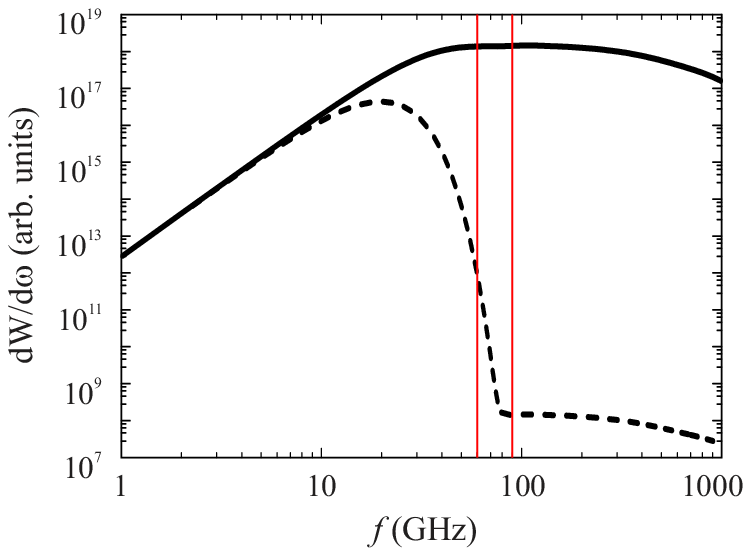}
\caption{CDR spectrum calculated for two bunch lengths of $3\,$mm (dashed line) and $0.03\,$mm (solid line). The red area shows SBD detector sensitivity range ($b = 50\,$mm, $a = 3\,$mm, $\gamma = 80$, $N = 10^{10}\,$e/bunch).}
\label{F3}
\end{figure}

Here $\theta_{max}=0.1\,$rad. Substituting Eq.\eqref{Eq5} in Eq.\eqref{GrindEQ__1_} one may calculate the CDR spectrum from the electron bunch of a certain duration. The integration in Eq.\eqref{Eq5} is performed over $0.1$rad angular acceptance. The result of the calculation is illustrated in Fig.\ref{F3}. The dashed line shows the spectrum generated by $3\,$mm RMS long bunch. The part on the left is a coherent emission, whereas the part on the right (above $100\,$GHz) is an incoherent one. It is clear that for a Gaussian longitudinal bunch profile the CDR intensity in the frequency range of SBD detector shown on the graph is rather low (close to the incoherent part). The THz frequency range radiation is purely incoherent and, therefore, the intensity is also low. 
Nevertheless, in linear accelerators longitudinal beam profile might seriously differ from the Gaussian one, i.e. the core of the bunch might emit radiation coherently at a much shorter wavelength. However the generation of THz radiation with such a bunch shape would be inefficient. Moreover, microbunch instabilities might cause a non-uniform longitudinal distribution as well as a non-stable radiation spectrum. To generate high intensity THz radiation it is necessary to make a much shorter electron bunch. The plans on upgrading the laser system with the one which has the pulse duration of the order of $100\,$fs (about $30\,\mu$m) are described below. In this case the CDR spectrum will expand into the THz region as shown in Fig.\ref{F3} (solid line).

The purpose of the current investigations was to study the properties of the open resonator at longer wavelengths where diffraction effect and other losses are large. If one will be able to observe the basic CDR characteristics, align the cavity, witness the stimulation process and measure the cavity quality factor in this wavelength range, certainly it will be possible to tune and align the cavity at shorter wavelengths. 
\section{Experimental setup}
\label{ES}
\subsection{KEK: LUCX facility}
\label{sES}
The experimental setup for the observation and investigation of the SCDR phenomenon has been designed and constructed at LUCX facility in KEK. The LUCX - is a multipurpose linear electron accelerator where two klystrons (Toshiba $E3729$ and Toshiba $E3712$) are used to independently feed normal conductivity RF Gun and accelerating structure \cite{15}. The electron beam parameters are summarized in Table \ref{T1}.

\begin{table}[htp]
\centering
\begin{tabular}{l|l}
\hline
Parameter                                               & Value \\
\hline \hline
Beam Energy                            & $43\,$MeV ($\gamma = 84$) \\
Intensity                              & $0.5\,$nC/bunch     \\
Number of bunches                      & $100$ \\
Bunch spacing					                 & $2.8\,$ns \\
RMS Bunch length                       & $<10\,$ps\\
Repetition rate, typ                   & $3.13\,$trains/sec  \\
Normalized horizontal emittance        & $5 \pi\,$mm mrad \\ 
Normalized vertical emittance          &  $4\pi\,$mm mrad \\
Beam size in SCDR chamber              & $200\,\mu$m$\times$$200\,\mu$m \\
\hline
\end{tabular}
\caption{LUCX beam parameters.}
\label{T1}
\end{table}

A new, standing wave $1.6$ cell large mode separation and a high Q-value RF Gun with emittance compensation solenoid is used to produce a multi-bunch high quality electron beam with up to $1000$ bunches, $0.5\,$nC bunch charge, and $5$MeV beam energy \cite{16,17}. To allow normal incidence illumination of $Cs_{2}Te$ cathode with $266\,$nm and $10\,\mu$J/pulse laser light, four dipole magnet chicane and laser transparent vacuum window have been implemented \cite{18}. In order to avoid space charge emittance growth in the subsequent beam line the beam is properly matched with a quadrupole triplet to the following $3\,$m S-band traveling wave accelerating structure \cite{17} where it is accelerated up to $43\,$MeV. After that it enters ``Compton chamber'' where it was used for X-ray generation \cite{19}. Then the electron beam is deflected by a bending magnet ($BH1$) and enters the SCDR open resonator. Figure \ref{F4} shows schematic of the LUCX beam line and SAD \cite{20} simulated electron beam optics used for SCDR experiment. 

LUCX electron beam diagnostics is represented by three subsystems: $10$ button type beam position monitors ($MB1G$, $MB2G$, etc.), $3$ inductive current transformers ($ICT1$, $ICT2$ and $ICT3$) and $4$ screen monitors ($MS2G$, $MS3G$, etc.). Signals from BPMs and ICTs are transmitted through $30$ meters high quality RF cables out of the accelerator tunnel to the DAQ based on $14$ bit CAMAC ADCs. Screen monitors use dual-position air actuators with installed luminescent screens and OTR (polished stainless steel) targets. $MS1G$ is the Faraday cup for absolute current measurements to tune the laser phase and RF gun phase. $MS3G$ screen monitor is combined with SCDR setup so the CCD camera captures OTR image produced from the first flat mirror of the microwave resonator when it is tilted by $45$ degree with respect to the beam trajectory. 

The image is acquired by $JAI$ $CV-M40$ analogue triggered CCD camera connected to a PCI frame grabber card. $MS5G$ screen is used for emittance measurements. All apparatus combined with a number of steering and quadrupole magnets, phase control of the Gun laser, accelerator RF and $3\,$m accelerating structure give the complete information and control of the electron beam. 

\begin{widetext}
\begin{center}
\begin{figure}[htbp!]
\centering
\includegraphics{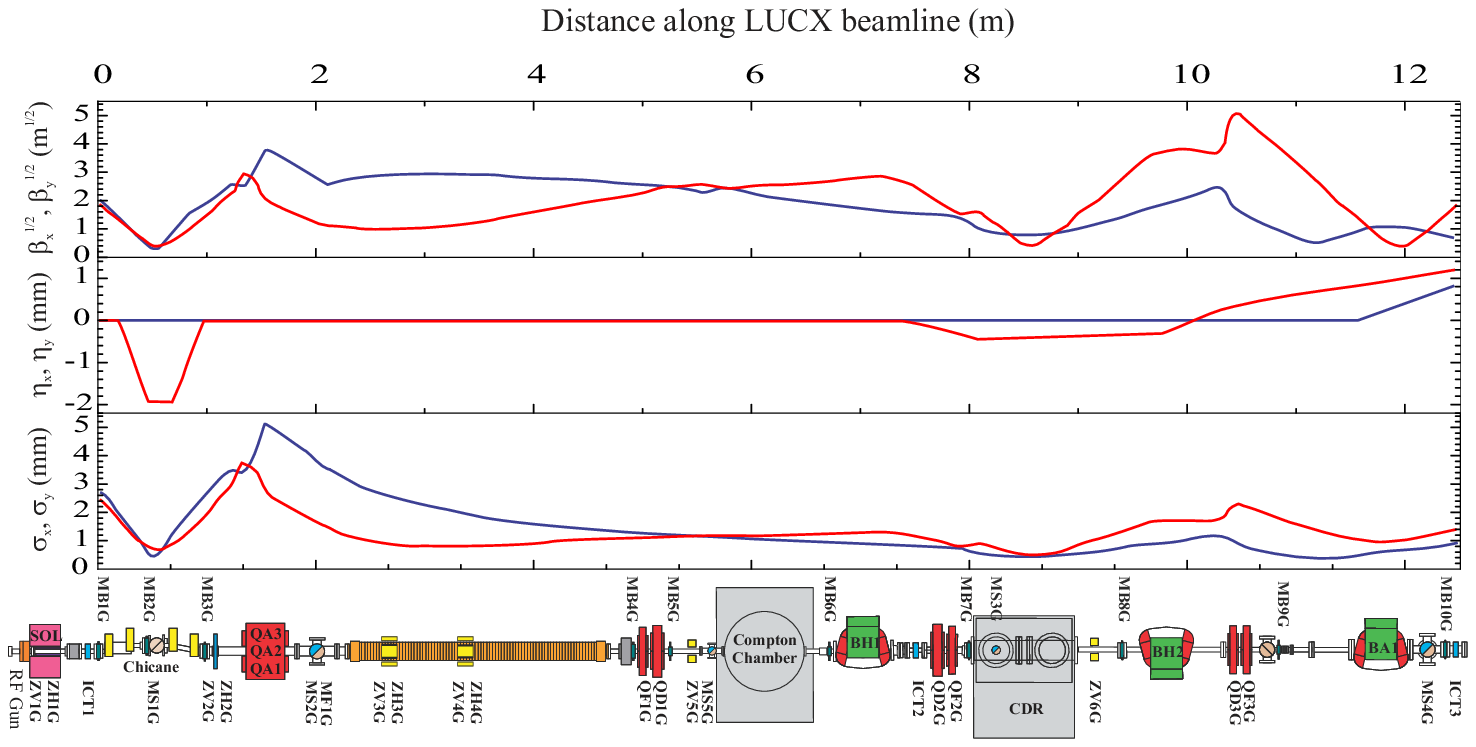}
\caption{From top to bottom: SAD simulated electron beam optics: beta function, dispersion and beam size along the beam line (red line - horizontal and blue line - vertical coordinates respectively); schematic of the LUCX beam line.}
\label{F4}
\end{figure}
\end{center}
\end{widetext}

\subsection{CDR resonator}

\begin{figure}[htbp!]
\includegraphics{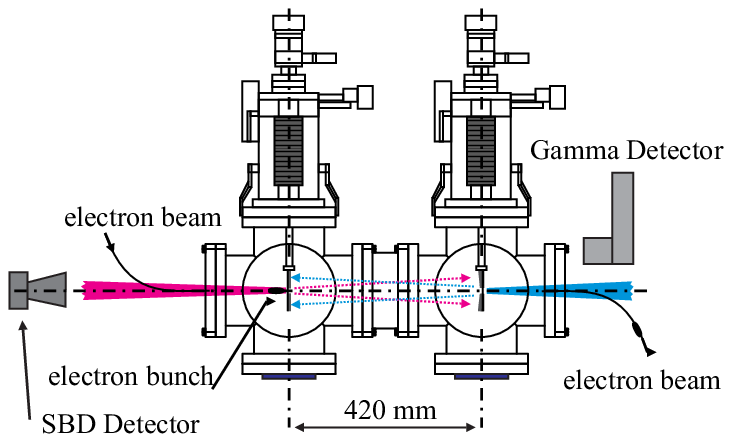}
\caption{Experimental layout.}
\label{F5}
\end{figure}

The experimental layout is shown in Fig.\ref{F5}. The installation design was well described in Refs. \cite{21,22}. In this section a brief overview of major design aspects is presented for convenience of the reader. The vacuum chamber consists of two $6-$way crosses connected by an extension part. The distance between the centers of the crosses was exactly half a distance between successive bunches in the train, i.e. $420\,$mm. The resonator mirrors were mounted on two identical $4$--Degree of Freedom (DOF) vacuum manipulation systems ($3$ translation DOF and $1$ rotation DOF). Another rotation DOF could be adjusted manually with adjustment screws of the in-vacuum mirror holders. Four DOF were motorized and remotely controlled. The first $100\,$mm diameter mirror was flat fused silica glass with a $5\,$mm free opening for the electron beam. The surface of the mirror was partially aluminized, i.e. there was a $15\,$mm hole in aluminum layer to extract a small fraction of the radiation out of the resonator for observation and tuning purposes, while keeping the rest of the radiation inside. The second $100\,$mm diameter mirror has a spherical concave surface with radius of curvature $840\,$mm (i.e. its focus was on the first mirror when resonator was aligned) and made of bulk aluminum with $5\,$mm free opening in its centre for free passage of the electron beam.

\subsection{Measurement system and data acquisition}

\begin{figure}[htbp!]
\includegraphics{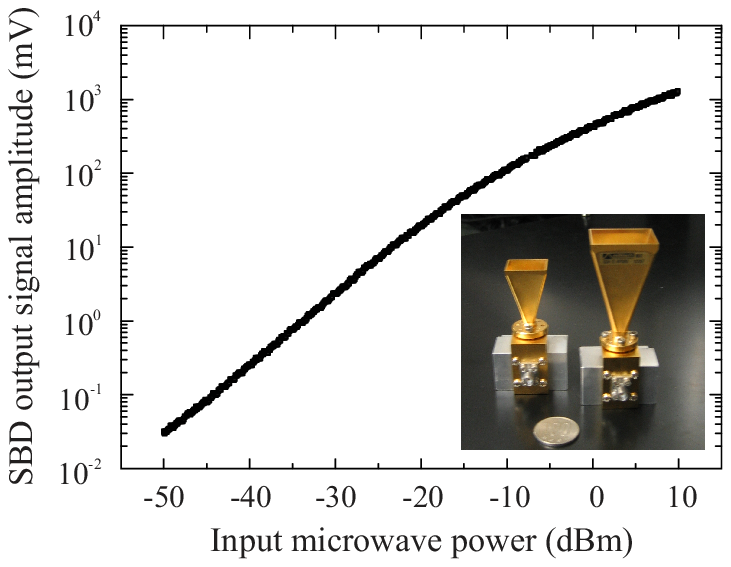}
\caption{Linearity of the SBD detector and photo of two SBD detectors (insert). The right detector is VDI DXP-12-RPFW0.}
\label{F6}
\end{figure}

The far-infrared radiation was observed upstream of the first mirror by an ultra-fast highly sensitive room-temperature Schottky Barrier Diode (SBD) detector with rated frequency response range of $60-90\,$GHz (see Fig.\ref{F6}). The performance of the detector was well described in Ref. \cite{23}. The SBD signal together with the ICT was acquired by a $1\,$GHz bandwidth, $5\,$GS/s Tektronix $684C$ Oscilloscope. The major characteristics of the detector are listed in Table \ref{T2}. 

\begin{table}[htp]
\centering
\begin{tabular}{l|l}
\hline
Parameter                                               & Value \\
\hline \hline
Frequency Range                    & $60 - 90\,$GHz \\
Wavelength range                   & $3.33 - 5\,$mm     \\
Response Time                      & $\sim250\,$ps \\
Antenna Gain					             & $24\,$dB \\
Input Aperture                     & $30 \times 23\,$mm\\
Video Sensitivity                  & $20\,$mV/mW  \\
\hline
\end{tabular}
\caption{Schottky Barrier Diode detector (VDI DXP-12) with Standard Gain Horn antenna.}
\label{T2}
\end{table}

When the open resonator is properly aligned the output radiation comes as a sequence of short pulses with $2.8\,$ns separation. The ultra-fast detector response allows for registering every pulse almost independently. It gives us an excellent opportunity to directly measure the radiation damping and the quality factor of the resonator.

\section{Experimental results and discussions}
\subsection{Accelerator tuning}

To generate a high-brightness electron beam at the LUCX facility the laser system based on a mode-locked $Nd:YVO_{4}$ oscillator and two flash-lamp-pumped amplifiers is used \cite{18}. This laser system requires daily maintenance that leads to a small position and angle variation of the laser beam on the cathode. The standard accelerator start-up includes adjustment of the horizontal and vertical laser beam positions and angles at the RF Gun photocathode with a remotely controlled mirror of the laser transport line. This procedure in combination with the electron beam position measurement at MB1G while varying solenoid field effectively represents the beam-based alignment. Also it optimizes the beam emittance and typically requires stability of both laser and accelerator RF systems for a high quality electron beam operation.  

The next step in tuning procedure is the RF Gun phase scan. The charge transmitted through the RF Gun is measured as a function of the relative phase between RF and the laser pulse. The laser injection phase can be set to minimize the energy spread of the beam coming out of the Gun down to $1\%$ RMS.  Afterwards the electron beam parameters and orbit are corrected by changing the RF phase of $3$ meter long accelerating structure, magnetic field of $BH1$, $BH2$ and $BA1$ bending magnets. Then two-dimensional emittance is measured at $MS5G$ OTR screen. The final emittance values are used in SAD code Ref. \cite{20} to calculate corrections for electron beam optics used for SCDR experiment. 

\begin{figure}[htbp]
\includegraphics{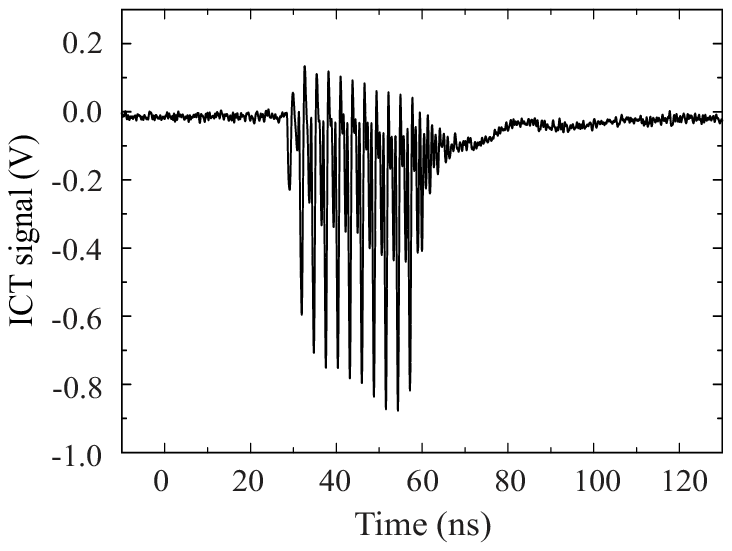}
\caption{Induction Current Transformer (ICT) signal.}
\label{F7}
\end{figure}

A typical ICT signal is illustrated in Fig.\ref{F7}. The bunch-by-bunch signal variation is fairly small and a slight growth of the signals from consequent bunches is the result of a low frequency noise from the high power RF system.

\subsection{CDR setup alignment and search for resonance}

\begin{center}
\begin{figure}[htbp!]
\includegraphics{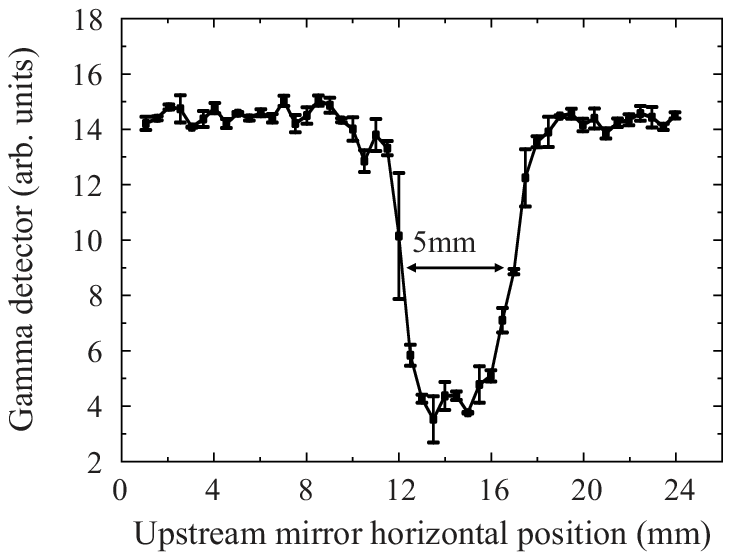}
\includegraphics{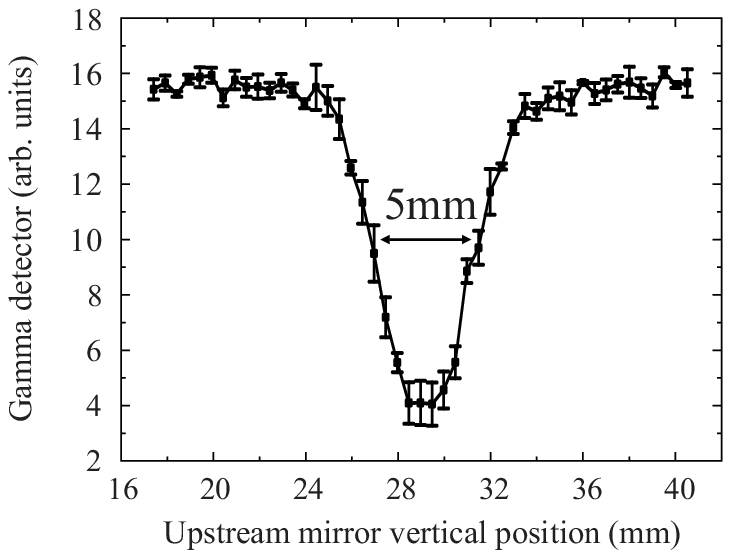}
\caption{Bremsstrahlung yield versus mirror position for horizontal (top) and vertical (bottom) upstream mirror displacement.}
\label{F8}
\end{figure}
\end{center}

Resonator mirrors were aligned with respect to the electron beam by measuring forward bremsstrahlung appearing due to direct interaction of the beam particles with the mirror material as a function of the mirror position. Examples of the scans are presented in Fig.\ref{F8}. The slopes of the dependencies are smeared out due to a non-zero beam dimensions. Moreover, the slopes are different for horizontal and vertical scans due to the difference in horizontal and vertical beam dimensions. The effective width of the bremsstrahlung signal corresponds to the mirror opening diameter. When the beam moves through the centre of the opening, where the bremsstrahlung dependence has a flat bottom, the signal is very small (as well as the number of particles interacting directly with the mirror material). These transverse positions of the resonator mirrors were used as initial set positions prior to resonator fine tuning. 

According to the estimations discussed in the theoretical part the SBD detector sensitivity range is close to the cut-off frequency for coherent radiation. On the other hand the incoherent radiation has a very low intensity and in the vicinity of the cut-off frequency microbunch instabilities might seriously affect the CDR shot-by-shot spectral content. As it was shown in \cite{22} the dependence of CDR power versus electron beam charge is non-linear which means that some part of the beam emits radiation coherently. Since the dependence is not quadratic, the longitudinal particle distribution is not Gaussian but might be highly asymmetric or have a very dense (but low population) core which generates radiation coherently and depends on the electron beam charge. 

\begin{center}
\begin{figure}[htbp!]
\includegraphics{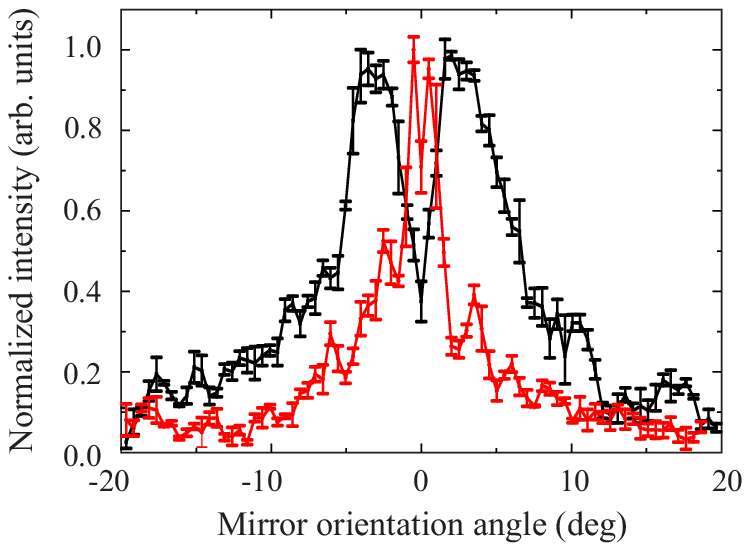}
\includegraphics{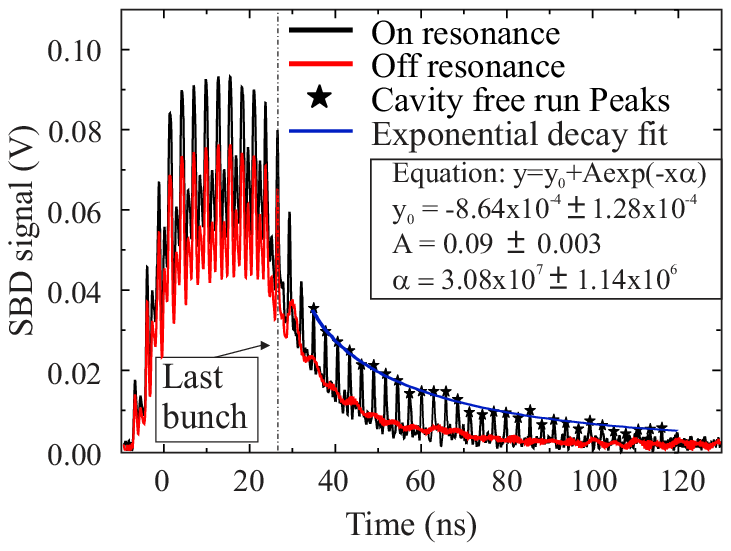}
\caption{Top: Orientation dependencies of the far-infrared signal versus mirror angle measured for the first (red line) and the second (black line) mirrors respectively. Bottom: Far-infrared signal from the SBD detector when the resonator is aligned (black line) and misaligned (red line) measured with $10$ bunches. The blue line represents an exponential fit for quality factor determination.}
\label{F9}
\end{figure}
\end{center}

The CDR distribution from a circular mirror can be represented as a combination of two identical polarization components with two-lobe distributions propagating in orthogonal directions. The SBD detector is polarization sensitive and detects only one polarization at a time. Figure \ref{F9} (top) illustrates the orientation dependencies of the far-infrared signal measured independently from each mirror. The widths of the distributions are different because the radii of curvature and distances from the mirrors to the detector are different. Moreover, the width of the distributions depends on the pre-wave zone effect contribution and diffraction through the output viewport.  

Nevertheless, it is very clear that both distributions have two peaks with a deep minimum in the centre. This is consistent with our expectations presented in the theoretical part. Knowing position of the minima one can perform initial alignment of the resonator mirrors with respect to each other and to electron beam to be able to setup the resonator optical axis.

In the experiment the resonator was set in such a way that the minima of the orientation dependencies from both mirrors coincide. When a bunch passes through the resonator it generates FDR from the first flat mirror. It propagates along with the beam and is reflected from the second concave mirror while the beam generates BDR from it. The concave second mirror focuses both the FDR and the BDR at the first mirror where they are partially reflected/transmitted through, while the subsequent bunch generates FDR. This cycle repeats itself for the duration of the train passing through the cavity. The transmitted part of the radiation was used to study the cavity and the radiation properties. The reflected part was stored in the resonator. If the distance between the mirrors was equal to a half of the bunch separation in the train, every following bunch generated radiation in the presence of a strong field from previous bunches leading to coherent radiation pulse stacking and the stimulation process. 

Finally one of the mirrors was moved back and forth along the electron beam propagation direction until clearer resonance was observed, i.e. the distance between the mirrors was exactly half the distance between the bunches ($420\,$mm or $1.4\,$ns). 
Figure \ref{F9} (bottom) shows a typical signal observed from the resonator when it is aligned (black line) and misaligned (red line) measured with $10$ bunches. In the beginning of the signal one can see a part when the beam was physically traveling through the vacuum chamber. It has a complicated structure caused by synchrotron radiation from upstream bending magnet $BH1$, diffraction radiation from upstream surface of the first mirror and radiation coming from the resonator itself. However, the length of the peaks structure is consistent with ICT signal shown in Fig.\ref{F7}. The tail of the signal (it is shown with dashed vertical line and starts after the spike of the coherent radiation from the last bunch in a train, i.e. when the beam train completely leaves the cavity) contains information about the radiation stored in the cavity only. That part of the signal was used to study the cavity properties. 

One may see that when resonator was aligned the detector registers every radiation round trip in the cavity. When the resonator was misaligned, i.e. its length is not exactly half the distance between the bunches the resonance disappears. To analyze the SBD signal tail behavior two methods were used. The first one was used for resonator alignment and was based on the analysis of the SBD signal tail peak integrals that corresponded to instantaneous CDR power stored in the cavity at a given round-trip. The second method was used to calculate the cavity quality factor. It includes the analysis of the peak amplitudes in the tail of SBD signal. The peak amplitude decay was characterized using an exponential function of the following form: 

\begin{equation} \label{GrindEQ__6_} 
A=A_{0} \exp \left[-\alpha t\right] 
\end{equation} 
\noindent where $t$ is the time and $\alpha$ is the decay constant. From this equation the cavity quality factor can be determined as:
\begin{equation} \label{GrindEQ__7_} 
Q = 2\pi\left|\frac{A}{\Delta A} \right|=2\pi\frac{1}{\alpha \Delta t}
\end{equation} 

\noindent with $\Delta A = \frac{dA}{dt} \Delta t=\alpha A_{0} \exp \left[-\alpha t\right]\Delta t$, where $\Delta t=2.8\,$ns is the resonator round-trip time. It is important to mention that the quality factor is independent of the initial intensity $A_{0}$. From the fit shown in Fig.\ref{F9} (bottom, blue line) the quality factor is equal to $72.88\pm3.8$. Using this decay-fitting technique it was possible to investigate resonator alignment accuracies and their effect on the quality factor.  

\begin{figure}[htbp!]
\includegraphics{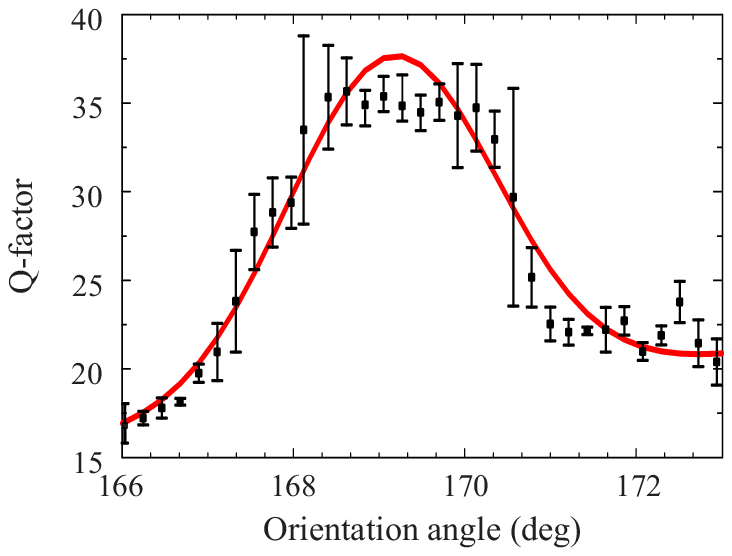}
\caption{Quality factor as a function of the orientation angle of the second mirror. The fit function Eq.\eqref{GrindEQ__8_} free parameters are: $y_{0} = -12.3$, $y_{1} = 0.09$, $y_{2} = 2.71$, $\overline{x} = 169\,$deg., and $\sigma = 1.22\pm0.05\,$deg.}
\label{F10}
\end{figure}

To show the resonator tuning sensitivity and to estimate the tolerance for the resonator mirror angular alignment we present a typical dependence of the quality factor as a function of the resonator mirror angle in Fig.\ref{F10}. The shape of the dependence is dictated by the cavity mirrors angular alignment and the distance between them. Figure \ref{F10} illustrates an example of the tuning procedure performed routinely during each experiment. As expected rotation of the mirrors results in misalignment of the resonator axis and degradation of the quality factor. The cavity alignment was performed with $2$ electron bunches in the train to avoid unnecessary background generation. However, the SBD signal tail was close to the Oscilloscope noise level. Since it was decided to exclude points less than $1\%$ above the noise level from being fitted to avoid non-physically large reconstructed quality factors. The maximum Q-factor in Fig.\ref{F10} is $35$ which is limited by the analysis accuracy. The distribution in Fig.\ref{F10} was fitted by the following function:

\begin{equation} \label{GrindEQ__8_} 
y(x)=y_{0} +y_{1} x+y_{2} \exp \left(-\frac{\left(x-\overline{x}\right)^{2} }{2\sigma ^{2} } \right) 
\end{equation} 

\noindent Here $y_{0}$, $y_{1}$, $y_{2}$, $\overline{x}$, and $\sigma$ are free parameters. The fit function is a sum of a linear function with a Gaussian. The reason for adding the linear component is due to the fact that when changing the orientation of the mirrors the radiation intensity changes (see Fig.\ref{F9}, top) together with the quality factor. As a result the pedestal in Fig.\ref{F10} also changes as a function of the mirror orientation. However, the $\sigma$ parameter is responsible for the width of the distribution and alignment accuracy. It is important to mention that the precision of the mirror angular adjustment is much smaller than the distribution width and, therefore, more than enough for cavity alignment. 

\subsection{Stimulation resonance study}

\begin{center}
\begin{figure}[hbp!]
\includegraphics{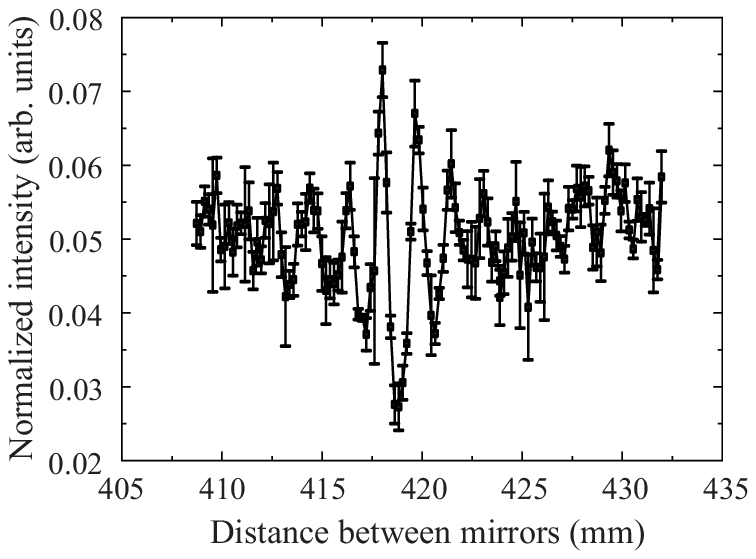}
\includegraphics{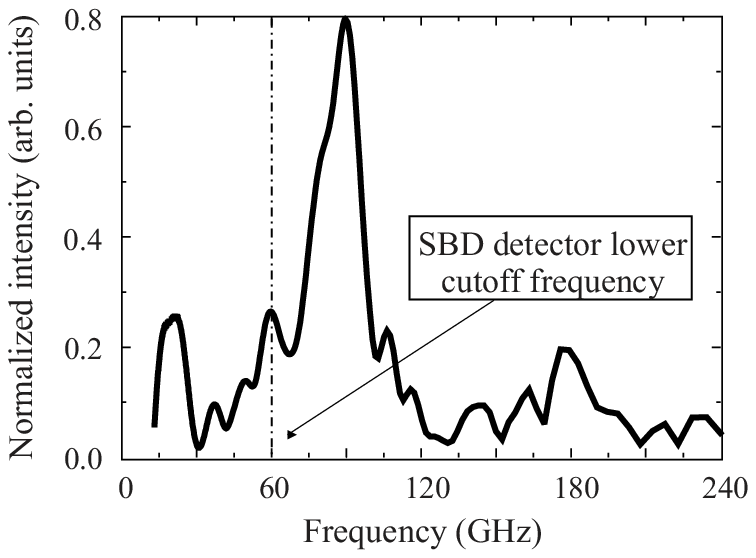}
\caption{Top: dependence of the radiation stored in the resonator (i.e. integrated over the signal tail) versus the cavity length measured with $10$ bunches; Bottom: FFT spectrum.}
\label{F11}
\end{figure}
\end{center}

To reveal more information about the CDR power stored in the cavity the dependence of microwave power on the cavity length was taken (Fig.\ref{F11}(top)). Effectively this resonator can be considered as a cavity with external ``pumping''. Variation of its length changes the pumping synchronicity or, in other words, the stored radiation resonance condition. By applying ``integral'' analysis technique to the acquired SBD signals the periodic interferogram function was observed. The periodicity appeared due to constructive and destructive interference of the stored and generated radiation while changing the cavity length. The envelope of the interferogram represents the CDR intensity enhancement due to coherent add up of the radiation pulses from different bunches in a train. The dependence was taken with $4$ electron bunches in the train. Therefore, only the cavity length, the CDR spectrum and the SBD spectral response contributed to the shape of the interferogram. The asymmetric shape of the interferogram can be explained by the fact that the scan takes time and the detector measures the CDR power at the edge where both coherent and incoherent processes contribute. In this case small shot-by-shot fluctuations of the longitudinal bunch shape might cause the signal variations. Nevertheless, the FFT spectrum (calculated according to \cite{24}) shown in Fig.\ref{F11} (bottom) demonstrates that its peak lays within the detector sensitivity range as expected.  However, the distribution is significantly different from the Fig.\ref{F3}. This aspect can be explained by the fact that the measurements were performed by a narrow band detector, i.e. on the right the cut-off is determined by the longitudinal bunch form factor, but on the left the cut-off is determined by the detector waveguide. To be able to perform detailed comparison with the theory, precise control over longitudinal bunch profile as well as a broadband detector of the same time response as SBD is needed. Nonetheless dependence at Fig.\ref{F11} was used to find an optimal cavity length corresponded to the maximum radiation gain. 

\begin{figure}[hbp!]
\includegraphics{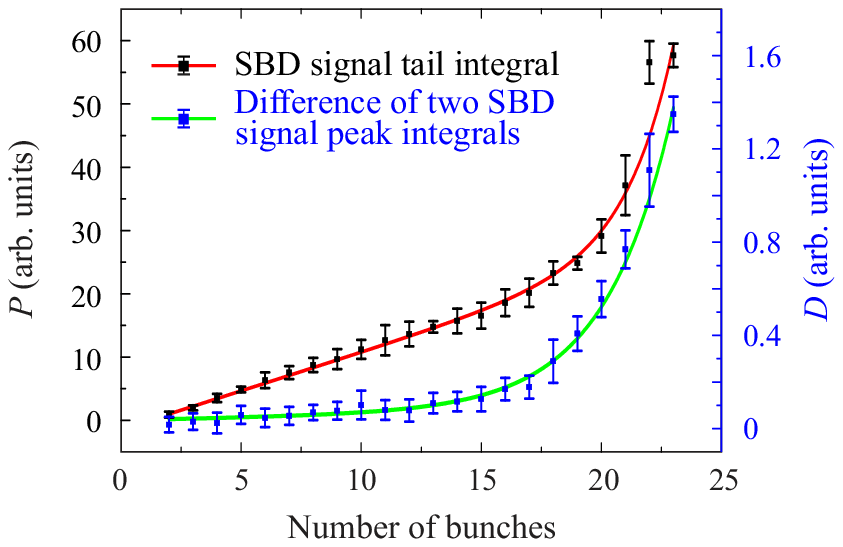}
\includegraphics{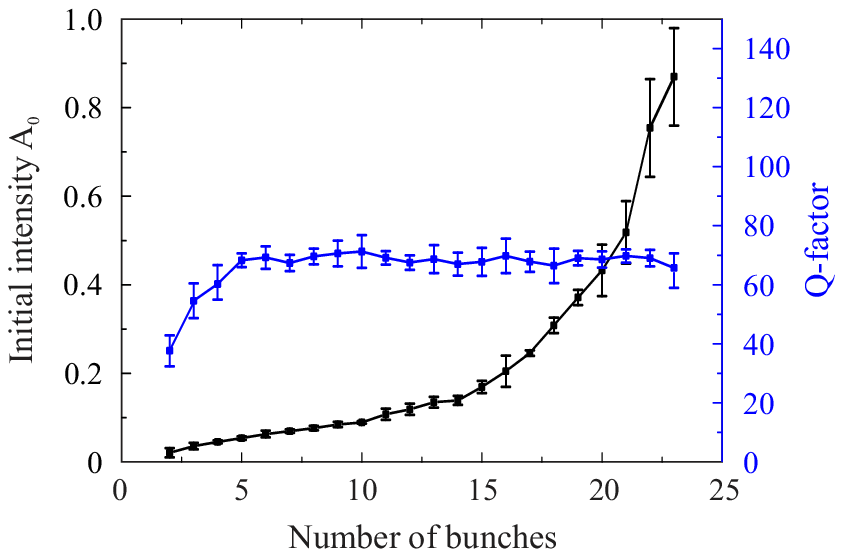}
\caption{Top: Dependence of the radiation power stored in the resonator as a function of the number of bunches in the train. SBD signal tail integral (black dots). Difference of two nearest SBD signal peak integrals (blue dots). The fit-function Eq.\eqref{GrindEQ__9_} parameters are: $y_{0} =  -1.44$, $y_{1} = 1.22$,$y_{2} = 2.5\times 10^{-4}$, $N_{0} = 1.96\pm0.28$ and $y_{0} =  0.035$, $y_{1} = 2\times 10^{-3}$,$y_{2} = 3.4\times 10^{-4}$, $N_{0} = 2.8\pm0.21$ for black and blue dots respectively. Bottom: Dependence of initial intensity $A_{0}$, Eq.\eqref{GrindEQ__6_} (black dots) and quality factor, Eq.\eqref{GrindEQ__7_} (blue dots) of the cavity for every acquisition on the number of bunches.}
\label{F12}
\end{figure}

To detect the presence of the stimulation process the number of electron bunches in the train was changed while keeping all geometrical resonator parameters the same. Figure \ref{F12} (top, black dots) shows the stored CDR power $P$ as a function of the number of bunches. One may see that the function is highly non-linear. The distribution was fitted by the following function:
\begin{equation} \label{GrindEQ__9_} 
y(N)=y_{0} +y_{1} N+y_{2} \exp \left(\frac{N}{N_{0} } \right) 
\end{equation} 
Here $y_{0}$, $y_{1}$, $y_{2}$, and $N_{0}$ are the free parameters. The fit confirms the fact that the stored radiation power grows non-linearly with the number of bunches. This effect is similar to the one appeared in undulators where the radiation photon yield grows up exponentially with the number of undulator elements. However, in undulators this dependence does not contain a clear linear part and tends to saturate when the number of periods reach a certain quantity. 

We believe that the linear component is determined by the shape of the detection system response. In \cite{23} it was demonstrated that even though the core of every SBD signal peak is very short  ($< 1$ns), a longer tail of the peak leads to the condition when the signal generated by a subsequent bunch sits on the tail from a preceding bunch. This fact is driving appearance of a linear component, which is dominant for a small number of bunches. The analysis have shown that this linear growth is identical for every round-trip detector signal and varies very slowly as a function of the signal time. To support this statement the difference of two nearest SBD round-trip signal peak intensities $D$ is plotted along with the Eq.\eqref{GrindEQ__9_} fit at the Fig.\ref{F12} (top, blue dots). It is clear that the linear component practically disappears while the exponential-like growth still exists. This dependence demonstrates the absolute energy loss from the resonator in one round-trip measured with an accuracy of the SBD detector as a function of the number of bunches. The energy losses are directly proportional to the energy stored in the cavity. 

Another important dependence can be found by re-analyzing the same data set in terms of the intensity decay of the SBD signal tail, which is described by an exponential function in Eq.\eqref{GrindEQ__6_} (see Fig.\ref{F12} (bottom)). This function was used to determine the initial intensity $A_{0}$ (black dots) and the quality factor (blue dots) of the cavity for every acquisition as a function of the number of bunches. The plot evidently shows that overall intensity of the accumulated CDR power non-linearly depends on number of bunches while the Q-factor remains almost constant. The reduction of the Q-factor for a small number of bunches is defined by the Oscilloscope noise level and the linear ramp in the intensity dependence originated from the detection system response (similar to Fig.\ref{F12}(top)). 

A pure coherent superposition of the radiation produced by different bunches would reveal a quadratic dependence as a function of a number of bunches, if the energy losses from the resonator were negligibly small. However, the quality factor demonstrates that we loose approximately $9\%$ of energy every round-trip at the best cavity alignment. It means that after 11 bunches we must observe some kind of saturation. Instead we observe an exponential growth (see Fig.\ref{F12}). For the aligned resonator the radiation photon yield increases due to an additional strong process. Since only the number of bunches changes it is evident that we observe \textit{stimulation} of the radiation emission. Also taking into account that the forward bremsstrahlung from the target is nearly zero the radiation observed is the \textit{Stimulated Coherent Diffraction Radiation}.

\subsection{Future improvements}

Undoubtedly further investigations with a larger number of bunches would be necessary to reach saturation of the SCDR process. However, in this case it is necessary to invest more time in understanding of the multibunch electron beam dynamics. A more careful electron beam tuning and control will be necessary in order to have a stable beam of predicted quality.

On the other hand, the replacement of current RF gun laser with a custom Ti:Sa laser system is under consideration. That will allow to generate electron bunches with duration smaller than $100\,$fs (about $30\mu$m) and intensity stability of the order of $< 1\%$ RMS. In this case the CDR spectrum will expand into the THz region. 

\section{Conclusion}

In this paper we presented the first observation and investigation of the Stimulated Coherent Diffraction Radiation (SCDR) in an open resonator as a possible mechanism for generation of intense THz radiation beams for numerous applications as a part of a larger program on THz sources development at KEK: LUCX facility. The use of an ultra-fast highly sensitive Schottky Barrier Diode detector allowed for a detailed investigation of the radiation stored in the resonator cavity as well as the cavity intrinsic properties. The turn-by-turn radiation damping and the resonator quality factor were measured. We clearly observed stimulation process as an exponential growth of radiation power stored in the resonator as a function of the number of bunches in the train. We assume that this exponential behavior is similar to the FELs gain factor as a function of the number of undulator periods. This fact suggests that there is a similarity between stimulation processes in undulators and open resonators.

It was demonstrated that the motion stability and mover accuracies are sufficient for cavity alignment and stable generation of SCDR beams. The exponential growth of the stored microwave radiation as a function of the number of bunches was observed. Despite the low resonator quality factor $72.88\pm3.8$ the SCDR process develops very rapidly and we do not observe any saturation even at $N=23$ bunches. Further investigations with larger number of bunches are required. However, it will also require more stable and well-tuned multibunch regime of the accelerator. This regime is currently under development and the results will be presented in a successive paper. Nevertheless in this article we presented an initial demonstration of a working prototype of a new pre-bunched beam pumped free electron maser based on SCDR process. If the power of the radiation is suitable for a third party experiments this system can provide a basis for future table-top high-brightness THz and X-ray sources. Also, the non-invasive nature of diffraction radiation enables to build a multi-user facility based on a linear accelerator by inserting a few cavities into the beam line. 

In the future we are planning to upgrade the RF gun laser system to be able to generate much shorter bunches and much higher power THz radiation beams. So far the basic characteristics of the generated and accumulated radiation were investigated and a good basis for future Thomson scattering experiment \cite{25} was made. In this new project the SCDR will be used as a radiation source.

\begin{acknowledgments}
The work was supported by the Quantum Beam Technology Program of the Japanese Ministry of Education, Culture, Sports, Science, and Technology (MEXT) and SASAKAWA foundation (Grant No. $3875$), JSPS KAKENHI Grant No. 23226020, and program ``Advancement of the scientific potential of high education'' by Russian Ministry of Education and Science No. $2.1799.2011$
\end{acknowledgments}


\begin{thebibliography}{}
\bibitem[1]{1} http://thznetwork.net/index.php/about/history

\bibitem[2]{2}  D. Dattoli, A. Renieri, and A. Torre, Lectures on Free Electron Laser Theory and Related Topics, World Scientific, 1993 

\bibitem[3]{3}  A. Gover and E. Dyumin, FEL Prize Lecture: Coherent Electron-Beam Radiation Sources and FELs: A Theoretical Overview, 2006 Proceedings of Free Electron Laser conference (FEL) \textbf{1}

\bibitem[4]{4}  H. Lihn, P. Kung, C. Settakorn, H. Wiedemann, D. Bocek, and M. Hernandez, Phys. Rev. Lett \textbf{76}, 4163 (1996).

\bibitem[5]{5}  Y. Shibata, K. Ishi, S. Ono, Y. Inoue, S. Sasaki, M. Ikezawa, T. Takahashi, T. Matsuyama, K. Kobayashi, Y. Fujita, E. Bessonov, Phys. Rev. Lett. \textbf{78}, 2740 (1997).

\bibitem[6]{6}  S. Sasaki, Y. Shibata, K. Ishi, T. Ohsaka, Y.Kondo, F. Hinode, T. Matsuyama, M.Oyamada, Nucl. Instrum. Methods Phys. Res., Sect. A \textbf{483}, 209 (2002).

\bibitem[7]{7}  Y. Shibata, S. Sasaki, and K. Ishi, Nucl. Instrum. Methods Phys. Res., Sect. A \textbf{483}, 440 (2002).

\bibitem[8]{8a}  V.L. Ginzburg, V.N. Tsytovich, Transition Radiation and Transition Scattering, Adam Hilger, Bristol, 1990

\bibitem[9]{8}  A. Potylitsyn, M. Ryazanov, M. Strikhanov, and A. Tishchenko, \textit{Diffraction Radiation from Relativistic Particles}, Springer, Berlin, 2010.

\bibitem[10]{9}  J. Nodvick and D. Saxon, Phys. Rev. \textbf{96}, 180 (1954).

\bibitem[11]{10}  CST Studio Suite, www.CST.com.

\bibitem[12]{11}  A. Potylitsyn, Nucl. Instrum. Methods Phys. Res., Sect. A \textbf{455}, 213 (2000).

\bibitem[13]{12}  P. Karataev, S. Araki, R. Hamatsu, H. Hayano, T. Muto, G. Naumenko, A. Potylitsyn, N.Terunuma, J. Urakawa, Nucl. Instrum. Methods Phys. Res., Sect. B \textbf{227}, 198 (2005).

\bibitem[14]{13}  P. Karataev, S. Araki, A. Aryshev, G. Naumenko, A. Potylitsyn, N. Terunuma, and J. Urakawa, Phys. Rev. ST Accel. Beams \textbf{11}, 032804 (2008) and references therein.

\bibitem[15]{14}  P. Karataev, Phys. Lett. A \textbf{345}, 428 (2005).

\bibitem[16]{15}  M. Fukuda, S. Araki, A. Deshpande, Y. Higashi, Y. Honda, K. Sakaue, N. Sasao, T. Taniguchi, N.Terunuma, J. Urakawa, Nucl. Instrum. Methods Phys. Res., Sect. A \textbf{637}, S67 (2011).

\bibitem[17]{16}  A. Deshpande, S. Araki, M. Fukuda, K. Sakaue , N. Terunuma, J. Urakawa, N. Sasao, M. Washio, Nucl. Instrum. Methods Phys. Res., Sect. A \textbf{600}, 361 (2009).

\bibitem[18]{17}  S. Liu, M. Fukuda, S. Araki, N. Terunuma, J.Urakawa, K. Hirano, N. Sasao, Nucl. Instrum. Methods Phys. Res., Sect. A \textbf{584}, 1 (2008).

\bibitem[19]{18}  K. Hirano, M. Fukuda, M. Takano, Y. Yamazaki, T. Muto, S. Araki, N. Terunuma, M. Kuriki, M.Akemoto, H. Hayano, J. Urakawa, Nucl. Instrum. Methods Phys. Res., Sect. A \textbf{560}, 233 (2006).

\bibitem[20]{19}  K. Sakaue, M. Washio, S. Araki, M. Fukuda, Y. Higashi, Y. Honda, T. Omori, T. Taniguchi, N.Terunuma, J. Urakawa, and N. Sasao, Rev. Sci. Instrum. \textbf{80}, 123304 (2009).

\bibitem[21]{20}  SAD -- Strategic Accelerator Design, http://acc-physics.kek.jp/SAD/ 

\bibitem[22]{21}  A. Aryshev, A. Araki, M. Fukuda, J. Urakawa, P. Karataev, G. Naumenko, A. Potylitsyn, L. Sukhikh, D. Verigin, K. Sakaue, in Proceedings of the 1st International Particle Accelerator Conference, Kyoto, Japan, 2010 (ICR, Kyoto, 2010), MOPEA053.

\bibitem[23]{22}  A. Aryshev, A. Araki, M. Fukuda, P. Karataev, G. Naumenko, A. Potylitsyn, K. Sakaue, L.Sukhikh, J. Urakawa, and D. Verigin, \textit{J. of Physics: Conf. Series} \textbf{236} 012009 (2010).

\bibitem[24]{23}  A. Aryshev, S. Araki, P. Karataev, T. Naito, N. Terunuma, and J. Urakawa, Nucl. Instrum. Methods Phys. Res., Sect. A \textbf{580}, 1544 (2007).

\bibitem[25]{24}  L. Frohlich, DESY-Thesis 2005-011

\bibitem[26]{25}  A.P. Potylitsyn, Phys. Rev. E \textbf{60}, 2272 (1999).
\end{thebibliography}
\end{document}